\documentclass[10pt,twocolumn,english,aps,prl,showpacs,superscriptaddress,groupaddress,floatfix,graphics,graphicx]{revtex4-1}

\usepackage[T1]{fontenc}
\usepackage[english]{babel}
\usepackage{amsmath,amssymb,bm}
\usepackage{graphicx}
\usepackage{xcolor}
\usepackage{multirow,makecell,float,placeins}
\usepackage[bookmarks,colorlinks,linkcolor=blue,citecolor=blue,urlcolor=blue]{hyperref}
\usepackage[all]{hypcap}
\usepackage{siunitx,nicefrac}
\usepackage{soul,orcidlink,xspace,bbold,comment}

\newcommand{\ket}[1]{| {#1} \rangle} 

\newcommand{\SAVBA}{\affiliation{Institute of Informatics, Slovak Academy of Sciences, 84507 Bratislava, Slovakia}}
\newcommand{\FFBG}{\affiliation{Faculty of Physics, University of Belgrade, 11001 Belgrade, Serbia}}
\newcommand{\UPJS}{\affiliation{Institute of Physics, Pavol Jozef \v{S}af\'{a}rik University in Ko\v{s}ice, 04001 Ko\v{s}ice, Slovakia}}
\newcommand{\SAVKE}{\affiliation{Institute of Experimental Physics, Slovak Academy of Sciences, 04001 Ko\v{s}ice, Slovakia}}
\newcommand{\NTC}{\affiliation{New Technologies Research Centre, University of West Bohemia, Univerzitní 8, CZ-301 00 Pilsen, Czech Republic}}
\newcommand{\FIT}{\affiliation{Faculty of Informatics and Information Technologies, Slovak University of Technology, 842 16 Bratislava 4, Slovakia}}

\begin{document}
\title{Machine learning protocol to identify pairing symmetries via quasiparticle interference imaging in Ising superconductors}

\author{Adam Hlo{\v z}n{\' y}} \SAVBA \FIT
\author{Jozef Hani{\v s}} \SAVKE
\author{Martin Gmitra} \SAVKE \UPJS \NTC
\author{Marko Milivojevi{\' c}} \SAVBA \FFBG
\date{\today}

\begin{abstract}
Identifying the pairing symmetry in unconventional superconductors is essential for reliably characterizing their superconducting states and for enabling their integration into realistic quantum devices. Here, we introduce a machine-learning-guided strategy to determine pairing symmetry from quasiparticle interference (QPI) data, which integrates first-principles calculations, tight-binding modeling, and symmetry-based classification of the superconducting pairing function. We demonstrate the approach on monolayer NbSe$_2$ as an experimentally accessible probe of superconductivity in real materials, within a single scalar-impurity Bogoliubov–de Gennes framework. Our analysis shows that the QPI-to-parameter inverse problem can be solved with high accuracy for most superconducting pairing channels in this setting, indicating that QPI carries rich, learnable information about the superconducting gap structure. Taken together, these results demonstrate that machine-learning-assisted QPI analysis provides a promising pathway for precise learning of superconducting pairing functions in quantum materials.
\end{abstract}
\maketitle

\section{Introduction}

The experimental identification of pairing symmetry in unconventional superconductors~\cite{S12,LYW+16,KWN+18}
is crucial for understanding the fundamental mechanisms behind superconductivity and for guiding the design of materials with novel quantum properties. This determination remains a central challenge in condensed matter physics due to the subtle and complex nature of the superconducting order parameter~\cite{FCS14,HKM11}. Techniques like Josephson junction measurements and momentum-resolved gap spectroscopy often face difficulties isolating intrinsic superconducting signals from material-specific scattering and quasiparticle interference effects~\cite{KT00,AMM+13,HNK+10}. This challenge is especially pronounced in more complex systems such as cuprates and iron-based superconductors, where the exact nature of the pairing symmetry is still debated~\cite{HKM11,J10}. Current proposals range from nodal d-wave to anisotropic s-wave. Addressing these controversies requires advanced, robust symmetry-informed methodologies capable of describing the momentum-dependent pairing functions that capture both the detailed k-space structure and the possible mixing of singlet and triplet components in the superconducting gap.  This refined understanding is essential for progressing toward the unambiguous identification of unconventional superconducting states.

The key to detecting the nature of superconductivity, and its topological character, lies in understanding the nature of the Cooper pairs, which are the fundamental building blocks of the superconducting state. This understanding can be achieved by analyzing quasiparticle interference (QPI), a phenomenon where quasiparticles scatter off defects or boundaries, resulting in spatial modulations of the local density of states that are captured in QPI spectra. While QPI patterns~\cite{HQS13,KCB+15,DGD+17,ROM+23,ZLW+24,CPW+25,ESM25}, measurable via techniques like scanning tunneling microscopy (STM)~\cite{CLE93,SPP+97,HBD+97}, can reveal signatures of superconducting pairing~\cite{LZH+13,KCL+22}, extracting and interpreting all the relevant information is a complex and challenging task~\cite{WZC+25}.  This complexity limits unambiguous identification of pairing symmetries in real materials and motivates usage of methods that are able to process large amounts of data.

\begin{figure*}[t]
    \centering \includegraphics[width=0.999\linewidth]{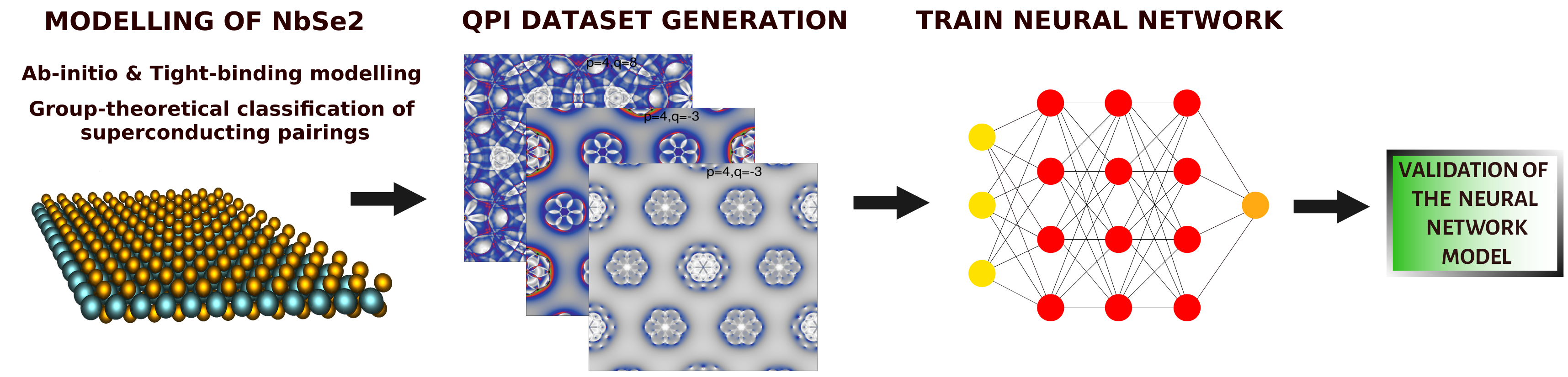}
    \caption{{\it Schematic overview of the proposed workflow.} Ab-initio calculations and tight-binding approach are used to model the normal phase of NbSe$_2$, whereas the superconducting pairing functions of NbSe$_2$ are determined using the group-theory approach. Secondly, a large dataset of simulated QPI spectra is generated across different superconducting parameters. This dataset is then used to train and optimize a neural network model.}
    \label{fig:enter-label}
\end{figure*}

Generating large datasets of QPI spectra through numerical simulations within the Bogoliubov–de Gennes (BdG) formalism offers a promising route to bridge the gap between theory and experiment. These simulations produce rich QPI patterns that capture subtle signatures of different superconducting pairing symmetries~\cite{HMG24}. 
However, the vast complexity and volume of these data demands advanced analytical tools. Machine learning approaches~\cite{LBH15}, such as convolutional neural networks (CNNs), have demonstrated powerful capabilities for pattern recognition~\cite{GWK+18} in images such as the QPI patterns, opening avenues for high-accuracy classification of pairing symmetries and extraction of hidden superconducting parameters from both simulated and experimental QPI spectra~\cite{ZMF+19}. Thus, combining symmetry‑based classification with machine learning‑assisted QPI analysis provides a transformative approach to uncovering the detailed nature of unconventional superconductivity in layered materials, in line with emerging machine‑learning methods for the study of superconductivity~\cite{TV23,SCC+25,LTL+26,NLK+26} and the extraction of microscopic parameters in quantum materials~\cite{SOT+23,KKL24,LL25}.

To address this, we introduce a machine-learning-guided scheme that integrates first-principles calculations, tight-binding modeling, and symmetry-based classification of superconducting pairing functions~\cite{VG85,KS19}, using QPI imaging as an experimentally accessible probe in real quantum materials. We demonstrate its usefulness on monolayer NbSe$_2$~\cite{CLA+22,Marganska2023:2DMat,SBA+25}, revealing how this machine learning-assisted QPI analysis provides a robust pathway for precisely determining pairing symmetries across diverse unconventional superconductors that can be described using the BdG formalism.

This paper is organized as follows. In Section~\ref{Description_of_superconductivity}, we 
provide theoretical framework for the analysis of  superconductivity parings in monolayer NbSe$_2$ based on BdG formalism, QPI imaging and machine learning. In Section~\ref{results}, we present our machine-learning-based results of subtracting superconducting pairing information directly from simulated QPI images of monolayer NbSe$_2$. Using a two-head CNN, we simultaneously classify the superconducting pairing type (classified according to the irreducible representation of the group of monolayer NbSe$_2$) and extract the parameters controlling the superconducting state, such as the singlet–triplet mixing and gap magnitude. Finally, in Section~\ref{conclusions}, we present our conclusions.

\section{Theoretical framework: superconductivity, QPI, and machine learning}\label{Description_of_superconductivity}
To fully describe and analyze superconductivity in 
NbSe$_2$ using quasiparticle interference imaging and machine learning methods, we first describe details of the monolayer NbSe$_2$ model in its normal state, followed by the description of possible supperconducting pairing within the single-band approximation of free-standing monolayer NbSe$_2$. As a second step, we provide the theoretical framework to generate QPI images based on scalar impurity, implementing different types of superconducting pairing function previously defined. Finally, we provide a machine-learning-based protocol, based on convolutional neural network,
able to analyze QPI images and solve the inverse problem of determining the superconducting pairing type and other microscopical details of the superconducting pairings.

\subsection{Bogoliubov-de Gennes description of superconductivity in monolayer NbSe$_2$}\label{BdGsubsection}

We are interested in studying the effects of electron superconducting pairing in free-standing monolayer NbSe$_2$, having the ${\bf D}_{3{\rm h}}$ symmetry, at a low-temperature regime. We model the superconductivity by the BdG formalism with the Hamiltonian 
\begin{equation}
    \mathcal{H}_\mathrm{BdG} = \frac{1}{2} \sum_\mathbf{k} \Psi^\dagger_\mathbf{k} \mathcal{H}_{\mathbf{k}}^\mathrm{BdG} \Psi_\mathbf{k},
\end{equation}
considering the Nambu spinor, $\Psi^\dagger_{\mathbf{k}} = [ c^\dagger_{\mathbf{k}\uparrow}, c^\dagger_{\mathbf{k}\downarrow}, c_{\mathbf{-k}\uparrow}, c_{\mathbf{-k}\downarrow} ]$ with fermionic creation $c^\dagger_{\mathbf{k}\uparrow}$ and annihilation $c_{-\mathbf{k}\uparrow}$ operators. The $4\times 4$ BdG Hamiltonian in reciprocal space $\mathcal{H}_{\mathbf{k}}^{\mathrm{BdG}}$ reads~\cite{LS18}
\begin{equation}\label{EQ:BDG}
    \mathcal{H}_{\mathbf{k}}^{\mathrm{BdG}} =
    \begin{pmatrix}
        H_\mathrm{e}(\mathbf{k}) & \Delta_{\mathbf{k}}\\
        \Delta^{\dag}_{\mathbf{k}}& -H_\mathrm{e}^{\mathrm T}(-\mathbf{k})
    \end{pmatrix},
\end{equation}
where the electron-like Hamiltonian $H_\mathrm{e}(\mathbf{k})$ is the effective tight-binding Hamiltonian 
\begin{equation}
    H_\mathrm{e}(\mathbf{k}) =  \mathcal{H}_\mathrm{orb}(\mathbf{k}) + \mathcal{H}_\mathrm{I}(\mathbf{k}),
\label{EQ:electron}
\end{equation}
In the equation above, the single-orbital part $\mathcal{H}_\mathrm{orb}$ describes the dispersion of the effective Nb $d$-band in the vicinity of the Fermi level~\cite{SM19,HMG24}
\begin{eqnarray}
&&{\cal H}_{\rm orb}({\bf k})=\varepsilon_0 + 2 t_1 f(\alpha,\beta)\\
&&+2 t_2 (\cos{2\beta}+2\cos{3\alpha}\cos{\beta})+2t_3 f(2\alpha,2\beta)\nonumber\\
&&+4t_4(\cos{\alpha}\cos{3\beta}+\cos{4\alpha}\cos{2\beta}+\cos{5\alpha}\cos{\beta})\nonumber\\
&&+2t_5 f(3\alpha,3\beta)+2 t_6 (\cos{2\beta}+2\cos{3\alpha}\cos{\beta})\nonumber\\
&&+4t_7(\cos{7\alpha}\cos{\beta}+\cos{5\alpha}\cos{3\beta} +
\cos{2\alpha}\cos{4\beta})\nonumber,
\end{eqnarray}
where  $f(\alpha,\beta)=\cos{2\alpha}+2\cos{\alpha}\cos{\beta}$, $\alpha=k_x a/2$, $\beta=\sqrt{3}k_y a/2$, $a$ is the lattice constant, $k_x$ and $k_y$ are components of the wave vectors in Cartesian frame, $\varepsilon_0$ is the chemical potential of the NbSe$_2$ monolayer, and $t_i$, $i=1,...,7$ are the real hopping parameters. The model describes Nb $d$-band close to the Fermi level up to 7-th neighbors.

In addition to the orbital part, one has to include the intrinsic spin-orbit coupling. The intrinsic spin-orbit coupling Hamiltonian $H_{\rm I}$~\cite{SM19} up to the third neighbor equals to
\begin{eqnarray}
{\cal H}_{\rm I}({\bf k})&=&2\sigma_z\Big(\lambda_{\rm I}^{(1)}g(\alpha,\beta)+\lambda_{\rm I}^{(3)}g(2\alpha,2\beta)\Big),
\end{eqnarray}
where $g(\alpha,\beta)=\sin{2\alpha}-2\sin{\alpha}\cos{\beta}$, $\sigma_z$ is the Pauli matrix, and the nonzero real parameters $\lambda_{\rm I}^{(1)}$ and $\lambda_{\rm I}^{(3)}$ quantify the strength of the interaction in the first and third neighbor approximation. The parameter $\lambda_{\rm I}^{(2)}=0$ due to vertical mirror plane symmetry.

Finally, $\Delta_{\mathbf{k}}$ represents ${\bf k}$-dependent superconducting order parameter.
Without entering into a macroscopic details of the superconductivity generation in a monolayer NbSe$_2$, it is possible to construct all allowed types of superconducting order parameters allowed by the ${\bf D}_{3{\rm h}}$ symmetry, divided into one singlet and two distinct triplet channels (parametrized by the spin pairing functions, $d_0\propto (\ket{\uparrow\downarrow}-\ket{\downarrow\uparrow})$ for singlet, and two triplet channels; one constructed by the $d_z\propto (\ket{\uparrow\downarrow}+\ket{\downarrow\uparrow})$ triplet and one by two spin $(d_x,d_y)$ $(d_x,d_y)\propto(\ket{\uparrow\uparrow}-\ket{\downarrow\downarrow},\ket{\uparrow\uparrow}+\ket{\downarrow\downarrow})$ and classified according to the irreducible representations (IRs) of the ${\bf D}_{3{\rm h}}$ group: one-dimensional (1D) IRs $A_1^u$, $A_1^g$, $A_2^u$, $A_2^g$, and two-dimensional (2D) IRs $E^u$ and $E^g$. In Appendix~\ref{appendixA}, we provide explicit forms of the superconducting gap functions classified according to IRs and the channel type: $(s)$ for a singlet, $(t,z)$ for a triplet constructed using the $d_z$ spinor solely, and $(t,xy)$ for a triplet constructed using $(d_x,d_y)$ multiplet.

It is physically reasonable to assume that a singlet-triplet mixing can occur since for each IR both the singlet and triplet functions can be realized. 
By taking into account that the triplet function $d_{xy}$ can not be mixed with the singlet and triplet function $d_z$ due to the opposite parity with respect to the horizontal mirror plane symmetry, we can obtain the following mixed gap functions that transform according to one-dimensional IRs
\begin{eqnarray}
\Delta_{A_1^u,{\bf k}}&=&\Delta\Delta_{A_1^u,{\bf k}}^{(t,xy)},\label{A1u}\\
\Delta_{A_2^u,{\bf k}}&=&\Delta\Delta_{A_2^u,{\bf k}}^{(t,xy)},\label{A2u}\\
\Delta_{A_1^g,{\bf k}}&=&\Delta(\cos{\theta}\Delta_{A_1^g,{\bf k}}^{(s)}+\sin{\theta}\Delta_{A_1^g,{\bf k}}^{(t,z)}),\label{A1g}\\
\Delta_{A_2^g,{\bf k}}&=&\Delta(\cos{\theta}\Delta_{A_2^g,{\bf k}}^{(s)}+\sin{\theta}\Delta_{A_2^g}^{(t,z)}),
\label{A2g}\end{eqnarray}
where $\Delta$ is the order parameter magnitude.
As one can notice, in the case of the 1D IRs $A_1^u$ and $A_2^u$ no mixing occurs, whereas in the case of 1D IRs $A_1^g$ and $A_2^g$ there is only one mixing parameter.

In the case of the 2D IRs the situation becomes more complicated since for each spin channel $c$, $c=(s), (t,z), (t,xy)$, gap function $\Delta_{E^{g/u},{\bf k}}$ that transforms according to the IR $E^{g/u}$ can be written as $(\cos{\theta_{c}}\Delta_{E^{g/u}_1,{\bf k}}^{c}+{\rm e}^{{\rm i}\varphi_{c}}\sin{\theta_{c}}\Delta_{E^{g/u}_2,{\bf k}}^{c})$, where $\Delta_{E^{g/u}_{1/2},{\bf k}}^{c}$ are the two (real) components of the 2D IR $E^{g/u}$ for spin channel $c$, $\theta_c$ represents the component mixing parameter, whereas $\varphi_c$ is time-symmetry breaking parameter. In addition to this, similar as in IRs $A_1^g$ and $A_2^g$, for IR $E^g$  there is a possibility of $(s)$ and $(t,z)$ spin channel mixing, giving rise to the most general form of gap functions (for more details, see Appendix~\ref{appendixA})
\begin{eqnarray}
\Delta_{E^u,{\bf k}}&=&\Delta(\cos{\theta_t}\Delta_{E^u_1,{\bf k}}^{(t,xy)}+{\rm e}^{{\rm i}\varphi_t}\sin{\theta_t}\Delta_{E^u_2,{\bf k}}^{(t,xy)}),\nonumber\\
\Delta_{E^g,{\bf k}}&=&\Delta\Big[\cos{\theta}
(\cos{\theta_s}\Delta_{E^g_1,{\bf k}}^{(s)}+{\rm e}^{{\rm i}\varphi_s}\sin{\theta_s}\Delta_{E^g_2,{\bf k}}^{(s)})\nonumber\\
&&+\sin{\theta}(\cos{\theta_t}\Delta_{E^g_1,{\bf k}}^{(t,z)}+{\rm e}^{{\rm i}\varphi_t}\sin{\theta_t}\Delta_{E^g_2,{\bf k}}^{(t,z)})\Big]\nonumber,
\end{eqnarray}
However, in generating our large QPI dataset, we neglect the influence of parameters $\varphi_{s/t}$ for the reasons explained in Sec.~\ref{results}. To keep the analysis conceptually simple and demonstrate that our machine-learning protocol can determine key microscopic parameters of the superconducting Hamiltonian, we restrict the $E^g$ representation to the singlet channel only. Therefore, the gap functions going to be used in the case of IRs $E^u$ and $E^g$ are 
\begin{eqnarray}
\Delta_{E^u,{\bf k}}&=&\Delta(\cos{\theta}\Delta_{E^u_1,{\bf k}}^{(t,xy)}+\sin{\theta}\Delta_{E^u_2,{\bf k}}^{(t,xy)})\label{2DEu},\\
\Delta_{E^g,{\bf k}}&=&\Delta(
\cos{\theta}\Delta_{E^g_1,{\bf k}}^{(s)}
+\sin{\theta}\Delta_{E^g_2,{\bf k}}^{(s)})\label{2Dsinglet},
\end{eqnarray}
where in both cases, for simplicity, the component mixing parameter is given as $\theta$. 

\subsection{Quasiparticle interference details}
QPI imaging, typically performed via Fourier-transform scanning tunneling spectroscopy, has emerged as a powerful tool for probing superconducting states in quantum materials, offering momentum-resolved insights into Bogoliubov quasiparticle scattering that complement angle-resolved photoemission spectroscopy. In superconductors, QPI patterns arise from interference between incoming and elastically backscattered quasiparticles off impurities or defects, with the local density of states (LDOS) modulation dispersing according to the joint density of states on the BdG quasiparticle spectrum.  Up to now, QPI has been positioned as a tool that can explain superconductivity in different systems~\cite{Byers1993,HML+02},
being able to identify superconducting gap features such as phase structure, sign \cite{Pereg-Barnea2003,Nummer2006} and orbital order \cite{Chen2023NatComm}.

Theoretical modeling of QPI in superconductors draws inspiration from the intuitive picture of a solid-state system as a sea of quasiparticles—wave-like excitations governed by the system's Hamiltonian. Small perturbations from impurities or crystal defects scatter these waves, creating standing wave modulations in the LDOS that scanning tunneling spectroscopy (STS) directly visualizes through conductance maps. In the superconducting state, these become Bogoliubov quasiparticle interferences, where elastic scattering mixes electron- and hole-like excitations, encoding the momentum-space structure of the superconducting gap.

The T-matrix formalism provides a framework to compute these patterns theoretically, starting from the bare Green's function of the unperturbed Hamiltonian and resumming scattering events off a localized impurity potential.
Since quasiparticle responses of many impurities possess identical poles as a single impurity case \cite{CSS03,ZAH04}, we can approximate the description of the elastic scattering process via a spin-conserving single impurity. Such impurity potential can be written in real Nambu space as $V = v \tau_3 \otimes \sigma_0$, where $v$ is the impurity strength, $\tau_3$ is the Pauli matrix for electron-holes and $\sigma_0$ is the identity operator in the spin space.
Considering the potential $V$ as a perturbation, the LDOS can be calculated using the T-matrix approach via Green's function written as
$G(\mathbf{k},\mathbf{k}'; \omega) = G_0(\mathbf{k},\mathbf{k}'; \omega)+ \delta G(\mathbf{k},\mathbf{k}'; \omega)
\label{eq_GF}$, where $G_0(\mathbf{k},\mathbf{k}'; \omega)$ is retarded bare Green's function. The correction $\delta G(\mathbf{k},\mathbf{k}'; \omega)$ can be found with the help of the
T-matrix, which solves the scattering problem for a single scalar impurity~ \cite{economou2006}. Here, $\omega = \varepsilon + \mathrm{i}\delta$ corresponds to
quasiparticle energy with small lifetime broadening ($\delta\approx 0.1$~meV). The spin-conserving scalar impurity $V$ leads to a T-matrix, $T(\omega) = V \cdot [\mathbb{1} - V \cdot G_0(\omega)]^{-1}$, where $G_0(\omega) = 1/\Omega \int [ \omega - \mathcal{H}_{\mathbf{k}}^{\rm BdG}]^{-1}$ is the integrated bare Green's function within the first Brillouin zone (BZ). The scattered quasiparticle amplitude is equal to 
\begin{equation}\label{density}
	\rho(\mathbf{q}; \omega) = -\frac{1}{\pi} \mathrm{Im} \left\{ \frac{1}{\Omega} \int \mathrm{d}\mathbf{k}\, \delta G(\mathbf{k}, \mathbf{k} + \mathbf{q}; \omega) \right\},
\end{equation}
in which the relation $\mathbf{k}' = \mathbf{k} + \mathbf{q}$ is used, with $\delta G$
integrated over all possible $\mathbf{k}$ points in the first BZ. The convolution theorem~\cite{kohsaka2017} enables efficient evaluation of QPI images $\rho(\mathbf{q}; \omega)$, reducing computational complexity and allowing analysis of fine-grid maps with minimal broadening.
Spectral characteristics of the band structure are calculated using the imaginary part of the spectral function (for more details, see~\cite{HMG24}).

\subsection{Machine-learning details}\label{MLdetails}
After preparing a dataset of simulated QPI images ($\sim$ 5000 train samples, $\sim$ 250 validation samples and $\sim$ 250 test samples, equal number of samples per IR, the final step is to train a convolutional neural network to solve the inverse problem: from a QPI pattern to the underlying parameters of the effective BdG description (IR label and continuous Hamiltonian parameters). We adopt a multi-task setting with two prediction heads: (i) a classification head that predicts the superconducting irreducible representation (IR), and (ii) a regression head that predicts the continuous parameters controlling the superconducting state.

Our primary model is based on a VGG-like architecture~\cite{SZ15,DZM+21}, see Appendix \ref{appendix:nn_arch} for details. Specifically, we use the convolutional \emph{features} block of VGG16 as a shared backbone and modify the input layer to accept single-channel QPI maps by replacing the first convolution with a $1\!\to\!64$ kernel of size $k\times k$ (default $k=3$), initialized with He/Kaiming normalization ~\cite{kaiming_normalization}. The resulting feature tensor is mapped to a fixed spatial size via adaptive average pooling to $7\times 7$, flattened, and passed through a fully connected aggregation layer,
\begin{equation}
\label{eq:agg_layer}
\mathrm{agg} = \mathrm{ReLU}(W\mathbf{x} + \mathbf{b}),
\end{equation}
where $\mathbf{x}$ is flattened output of the convolutional feature extractor, $W$ is weight matrix of the layer, $\mathbf{b}$ is bias term and ReLU is defined as:
\begin{equation}
\label{eq:relu}
    \mathrm{ReLU}(t) = \max(0, t).
\end{equation}
From this shared representation, two parallel heads are evaluated:
(i) a classification head producing a score/probability vector over IRs, and (ii) a regression head producing four continuous outputs.
In addition to this VGG-based model, we implemented an analogous modification of ConvNeXt ~\cite{liu2022convnet}; however, for the present task and dataset size, the simpler VGG-based backbone provided more stable training and similar or marginally better overall validation performance.
\begin{figure*}[t]
    \centering \includegraphics[width=0.75\linewidth]{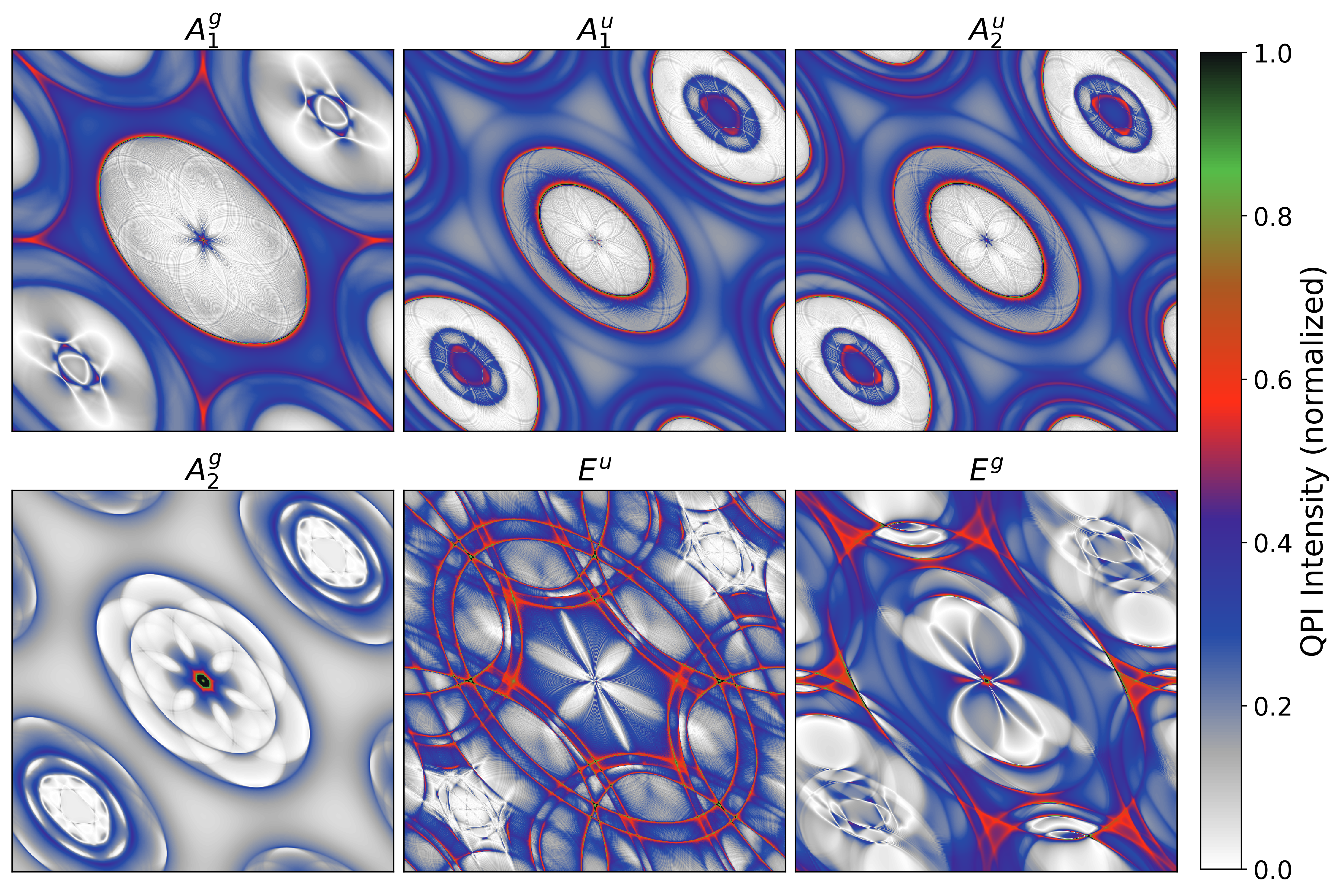}
\caption{Representative QPI images for each IR $A_1^g$, $A_1^u$, $A_2^u$, $A_2^g$, $E^u$, and $E^g$ within the 
validation dataset.}\label{fig:QPI}
\end{figure*}

The classification target is the IR label of the superconducting order parameter used to generate the QPI image.
The regression target consists of the chemical potential $\mu$, superconducting gap magnitude $\Delta$, and the mixing angle $\theta$ entering gap functions in Eqs.~\eqref{A1g}-\eqref{2Dsinglet}.
To avoid discontinuities associated with the periodicity of the mixing angle, we do not predict $\theta$ directly. Instead, we predict the two-component embedding
\begin{equation}
(\theta_1,\theta_2) = \left(\sin 2\theta,\;\cos 2\theta\right),
\end{equation}
which correctly enforces the physical periodicity $\theta \equiv \theta+\pi$ in our problem (see the discussion in the next Section for further explanation).
At inference, $\theta$ is reconstructed by $\theta = \tfrac{1}{2}\mathrm{atan2}(\theta_1,\theta_2)$ (modulo $\pi$), where atan2 is arc tangent of $\theta_1 /\theta_2$ with proper choice of quadrant.
The model is trained with a weighted multi-task objective function,
\begin{equation}
\mathcal{L} \;=\; w_{\mathrm{IR}}\,\mathcal{L}_{\mathrm{CE}}\!\left(\hat{\mathbf{p}}, \mathbf{y}_{\mathrm{IR}}\right)
\;+\;
w_{\mathrm{param}}\,\|\hat{\mathbf{r}}-\mathbf{y}_{\mathrm{param}}\|^2,
\label{eq:twoloss}
\end{equation}
where $\hat{\mathbf{p}}$ denotes the predicted IR scores/probabilities, $\mathbf{y}_{\mathrm{IR}}$ one-hot encoded IR ground truths,  $\hat{\mathbf{r}}=(\hat{\theta}_1,\hat{\theta}_2,\hat{\mu},\hat{\Delta})$ predicted parameters of the BdG Hamiltonian, and $\mathbf{y}_{\mathrm{param}}$ the corresponding ground-truth targets. Furthermore, $w_{\mathrm{IR}}$, $w_{\mathrm{param}}$ are weights of the IR and BdG parameter parts of the objective function, with $\mathcal{L}_{\mathrm{CE}}$ denoting cathegorical cross-entropy defined as:
\begin{equation}
\mathcal{L}_{\mathrm{CE}}\!\left(\hat{\mathbf{p}}, \mathbf{y}_{\mathrm{IR}}\right)
=
- \sum_{i=1}^{K} \left( \mathbf{y}_{\mathrm{IR}} \right)_i \log \hat{p}_i,
\end{equation}
with summation running across possible IRs. The loss $\mathcal{L}$ described in Eq.\ref{eq:twoloss} is per-sample, the final loss depends on the batch size and is calculated as average across all samples.

As a way to regularize the training dataset and reduce overfitting ~\cite{bishop_noise_augmentations}, we apply mild noise augmentations to the input QPI images during training - Gaussian noise and pink (1/f) noise proved to slightly improve stability of the training and validation accuracy.

\section{Results}\label{results}
A large set of QPI images, with representative examples for each IR shown in Fig.~\ref{fig:QPI}, were obtained for different set of superconducting pairing functions and chemical potentials $\varepsilon_0$, while keeping the same values
of the hopping parameters $t_1,...,t_7$ (in meV) and
intrinsic spin-orbit coupling parameters $\lambda_{\rm I}^{(1)}$ and $\lambda_{\rm I}^{(3)}$ (also in meV) to
$t_1=33.52$,
$t_2=97.26$, 
$t_3=-2.11$,
$t_4=-13.53$, 
$t_5=-10.30$,
$t_6=3.48$, 
$t_7=1.69$,
$\lambda_{\rm I}^{(1)}=13.27$, and
$\lambda_{\rm I}^{(3)}=-1.94$, respectivelly~\cite{HMG24}. On the other hand, we have varied the chemical potential $\varepsilon_0$ values, to test whether this framework is applicable to a class of TMDC-based misfit structures~\cite{RMR+93,NLB+98}, which, despite their more complex layering, possess a band structure near the Fermi level that closely resembles that of the monolayer TMDCs, albeit with a modified chemical potential resulting from electron doping~\cite{LPC+20}. The above set of parameters fully describe the electron and the hole part of the BdG Hamiltonian; the final step is to define the allowed set of the superconducting parameters
for superconducting gaps defined in Eqs.~\ref{A1u}-\ref{2Dsinglet}.
Additionally, for all gaps in Eqs.~\ref{A1u}-\ref{2Dsinglet}, we have analyzed the effect of the superconducting gap intensity $\Delta$. At the same time, this is the only free parameter
to be varried in the case of superconducting gap functions that transform according to IRs $A_1^u$ (\ref{A1u}) and $A_2^u$ (\ref{A2u}), whereas in the case of gaps $\Delta_{A_1^g}$ (\ref{A1g}) and  $\Delta_{A_2^g}$ (\ref{A2g}) there is an additional parameter $\theta\in[0,2\pi]$ quantifying the level of singlet-triplet mixing, which was randomly sampled.
Finally, we move to gap transforming to 2D IRs $E^g$ and $E^u$, given in Eqs.~\eqref{2DEu} and~\eqref{2Dsinglet}. In the two case studied, besides varying the value of $\Delta$, it is allowed to vary parameter $\theta$, quantifying the mixing between two components of 2D IR $E^{g/u}$. 

Before performing the neural network analysis, we
tried to simplify the analysis by understanding the influence of free superconducting parameters on QPI images. Secondly, we note that QPI images are insensitive on the sign of the gap function, sugessting that, independently on the IR $\mu$, gaps $\Delta_{\mu,{\bf k}}$ and $-\Delta_{\mu,{\bf k}}$ are indistiguishable. Taking this into the account, it is clear that we can restrict ourselves to a resticted range of allowed superconducting parameters: positive $\Delta$ and $\theta$, ranging from 0 to $\pi$, instead from 0 to $2\pi$, since $\theta\rightarrow\theta+\pi$ change induces a sign change of $\Delta_{\mu,{\bf k}}$, which is indistinguishable by QPI imaging.
Additionally, our analysis of the QPI patterns indicates that it is not possible to properly determine $\varphi (\varphi_s)$ using (time-reversal conserving) scalar impurities. Preciselly this fact motivated us to use Eqs.~\eqref{2DEu} and~\eqref{2Dsinglet} in which $\varphi (\varphi_s)$ is neglected and suggests that (time-reversal breaking) magnetic impurities should be employed instead. We leave this investigation for future work.

\begin{table}[t]
\centering
\caption{{\it Evaluation metrics for detecting superconducting pairings.} For each IR (with $A_1^u$ and $A_2^u$ merged into a single class $A_{12}^u$ as explained in the main text), we report the percentage probability of correctly identifying the superconducting pairing transforming according to each IR, followed by the mean absolute errors for the continuous parameters $\Delta$, $\theta$, and $\varepsilon_0$ of the BdG Hamiltonian~\eqref{EQ:BDG}.}
\label{tab:ml_results}
\begin{tabular}{lcccc}
\hline\hline
IR &
hit $[\%]$ &
\makecell{$\delta\Delta$ [meV]} &
\makecell{$\delta\theta$} [rad]  &
\makecell{$\delta\varepsilon_0$ [eV]} 
 \\
\hline
$A_{1}^g$  & 86  & 0.01 & 0.30 & 0.07 \\
$A_{12}^u$ & 94 & 0.01 & /    & 0.01 \\
$A_{2}^g$  & 98 & 0.01 & 0.06 & 0.02 \\
$E^{g}$   & 98 & 0.01 & 0.01 & 0.01 \\
$E^{u}$   & 96  & 0.01 & 0.05 & 0.01 \\
\hline\hline
\end{tabular}
\end{table}

Finally, we notice that the QPI signatures for $\Delta_{A_{1}^{u},{\bf k}}$ and $\Delta_{A_2^{u},{\bf k}}$ become indistinguishable for the same underlying parameters in our setup. This can be traced back to the fact that, within the present modeling, the impurity effectively probes a gap opening predominantly at the Fermi level (we used $\omega=0$ in Eq.~\ref{density}) rather than symmetrically above/below it (nonzero $\omega$), where the differences between gap functions $\Delta_{A_{1}^{u},{\bf k}}$ and $\Delta_{A_{2}^{u},{\bf k}}$ are the most prononced. We therefore merge $A_{1u}$ and $A_{2u}$ into a single class during training and evaluation (we will call this class $A_{12u}$ and leave the analysis of different $\omega$ for future work(s). 

We now turn to the results of the machine-learning based analysis of the QPI data. As discussed in Section~\ref{MLdetails}, we evaluate the trained network by measuring (i) the representation hit rate, defined as the fraction of samples with a given ground-truth IR for which the predicted IR (argmax over $\hat{\mathbf{p}}$) matches the label, and (ii) the mean absolute error (MAE) for each regressed parameter within each ground-truth IR subset. For the mixing angle, the MAE is computed after reconstructing $\theta$ from $(\sin 2\theta,\cos 2\theta)$, respecting the $\pi$-periodicity.

Table~\ref{tab:ml_results} summarizes the per-representation performance metrics for detecting superconducting pairings, including classification probabilities and MAE for $\Delta$, $\theta$, and $\mu$. Overall, the model achieves near-perfect recognition for most IRs, while the reduced hit rate for $A_1^g$ and $E^u$ reflects the intrinsically weaker and/or more ambiguous QPI signatures available to a scalar impurity in these cases. 
This represents the main result of our work: machine learning can detect different superconducting pairing symmetries with very high success rates, even using scalar impurities and a single excitation energy ($\omega=0$), paving the way for precise identification of pairing functions in quantum materials like monolayer NbSe$_2$. 

A second important result is that the trained neural network is able to recognize superconducting parameters of the BdG Hamiltonian~\cite{KKL24}: the regression errors $\delta\Delta$ remain small for the gap magnitude $\Delta$ across all IRs, indicating that $\Delta$ leaves a robust imprint on the QPI patterns. Additionally, once the $\theta\equiv\theta+\pi$ ambiguity is resolved via the $(\sin{2\theta}, \cos{2\theta})$ embedding, the trained network reliably determines the presence of the singlet-triplet (component) mixing; however, the precision of obtaining the exact $\theta$ parameter differs among different IRs. The same IR dependence on the precision of the chemical potential value is evident from the results in Table~\ref{tab:ml_results}; nonetheless, these values can be meaningfully compared to those extracted via first-principles calculations combined with tight-binding modeling.

We note that orbital-dependent features in the QPI spectrum have been explicitly identified in related systems such as 3R-NbS$_2$~\cite{Machida2017}, where the valley contrasting orbital texture leads to strongly asymmetric scattering channels. In monolayer NbSe$_2$, the experimental signatures of superconductivity, including the large in-plane critical field, are well captured within the single-band framework~\cite{CLA+22}. Therefore, our single-band model is appropriate for the present study. More generally, our workflow is naturally extensible to multiband superconductivity: if orbital-dependent features are present in the experimental QPI data of a given material, the tight-binding model and the QPI simulation pipeline can be straightforwardly extended to include multiple orbitals and the corresponding Wannier projections~\cite{Kreisel2015, Naritsuka2023}, increasing the fidelity of the theoretical model to experimental data.

Furthermore, in Ising superconductors with critical temperatures below 10~K, experiments are typically performed at temperatures well below $T_c$~\cite{CLA+22}, where $k_BT \ll \Delta$ and thermal broadening does not significantly affect the QPI features. For materials with higher critical temperatures, finite-temperature broadening can be incorporated into the simulation pipeline by generating temperature-modified QPI datasets and retraining the network accordingly.

\begin{figure*}[t]
    \centering \includegraphics[width=0.999\linewidth]{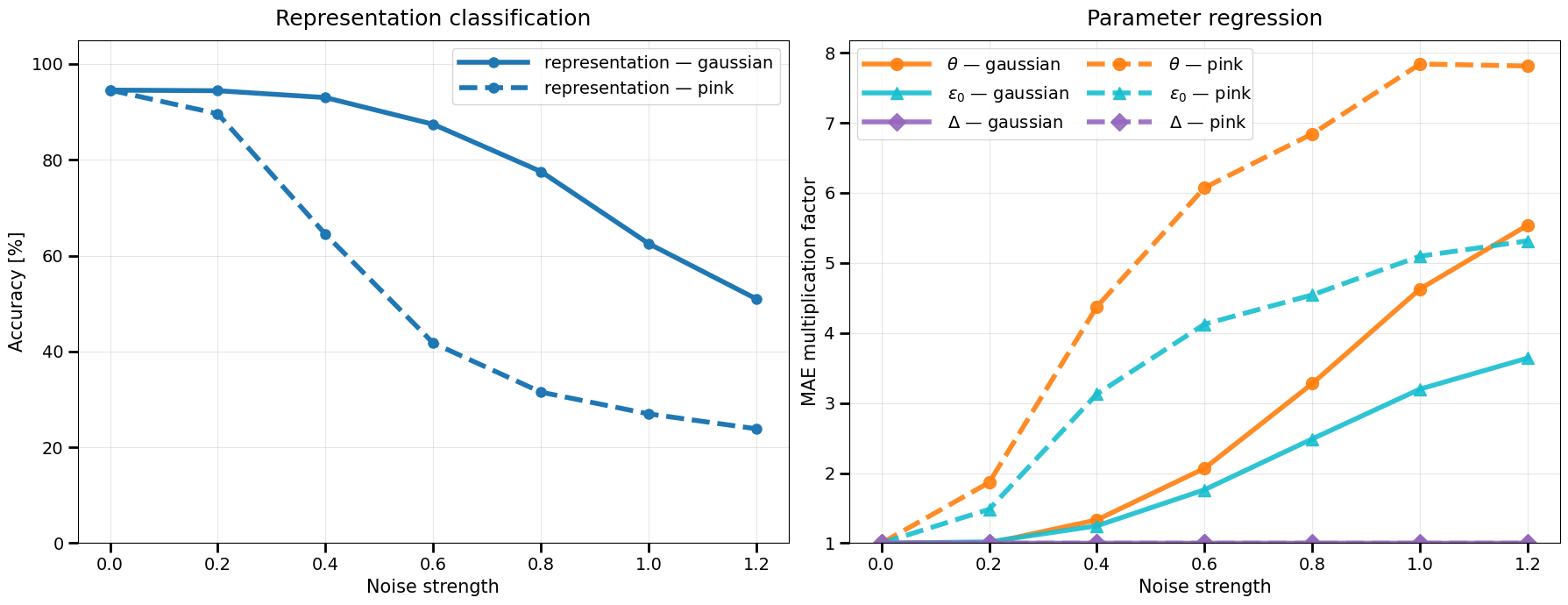}
    \caption{{\it Robustness of the model to the added noise}. The left plot shows representation classification accuracy under Gaussian and pink $(1/f)$ noise. The right plot shows regression degradation for $\theta$, $\varepsilon_0$, and $\Delta$, plotted as the ratio between the MAE at a given noise level and the MAE at zero noise. The noise strength is the noise standard deviation relative to the input image standard deviation.
}\label{fig:noise}
\end{figure*}

The robustness of the trained model to experimental noise deserves brief discussion. To bridge the gap between simulations and real experimental data which contains thermal, electrical, and speckle noise, we augmented the training dataset with Gaussian and pink (1/f) noise. The pink noise was chosen deliberately, as its power-law spectral profile mimics the low-frequency fluctuations characteristic of STM measurements. Additionally, since the network sees a different noisy realization of each QPI image at every training step, it cannot memorize specific training samples and it is  forced to learn features that are stable across noise (physically meaningful features of the QPI patterns), which is reflected in the improved validation accuracy that we observed after its introduction~\cite{bishop_noise_augmentations}.

To quantify the robustness of the trained model, we evaluated its performance on test images influenced by Gaussian and pink 
noise of increasing strength, defined as $\sigma_{\text{noise}}/\sigma_{\text{image}}$, where $\sigma_{\text{noise}}$ 
is the standard deviation of the added noise and $\sigma_{\text{image}}$ is the standard deviation of the clean QPI image. The results are shown in Fig.~\ref{fig:noise}. The left panel shows that the representation classification accuracy remains above 85\% for Gaussian noise up to strength $\sim 0.6$, indicating that the model is robust under moderate noise, where noise fluctuations are 60\% of the typical signal fluctuations. The network is more sensitive to pink noise than to Gaussian noise, 
which can be understood from the fact that pink noise produces spatially extended, slowly varying distortions of the QPI image that are more difficult for the network to separate from the physical signal than the pixel-scale fluctuations of Gaussian noise.

The right panel of Fig.~\ref{fig:noise} shows the regression for the continuous parameters $\theta$, $\varepsilon_0$, and $\Delta$, quantified using the MAE multiplication factor that represents the ratio of the MAE at a given noise level to the MAE at zero noise. The gap magnitude $\Delta$ remains essentially unaffected across the full range of noise strengths tested, confirming that $\Delta$ leaves a robust imprint on the QPI patterns, consistent with the uniformly small regression errors $\delta\Delta \approx 0.01$~meV reported in Table~\ref{tab:ml_results}. In contrast, the mixing angle $\theta$ is the most sensitive parameter, with its MAE increasing by a factor of up to $\sim 8$ at the highest noise level tested (noise strength 1.2), followed by the 
chemical potential $\varepsilon_0$. This establishes a clear hierarchy of the robustness of the trained features of our neural network: the symmetry classification (IR label) and the gap magnitude $\Delta$ are the most robust under noise, while the mixing angle $\theta$ is the first to degrade when increasing noise, consistent with the stronger IR dependence of $\delta\theta$ already observed in the noiseless case in Table~\ref{tab:ml_results}.

Looking forward, two complementary strategies exist for applying our framework to experimental data. If the dominant noise sources of a specific setup are known, physics-informed augmentation can be refined accordingly. Alternatively, self-supervised denoising can be applied as a preprocessing step directly to experimental QPI images, without requiring clean data. Kuijf et al.~\cite{KTB+25} recently demonstrated this approach for QPI data using the Noise2Self algorithm, outperforming standard Gaussian filtering while preserving fine-scale features, validated on both simulated and experimental cuprate data. These two strategies, used in combination, provide a concrete route toward applying our machine-learning 
protocol to real experimental measurements.

We also briefly address the case of multiple identical impurities. Since the poles of the multi-impurity Green's function, which encode the pairing symmetry, are identical to those of the single-impurity case~\cite{CSS03, ZAH04}, single-impurity approximation is justified. The multi-impurity case therefore does not remove the symmetry-encoding features of the QPI pattern; it introduces an additional amplitude modulation arising from the spatial distribution of the impurities. In the case of a small number of well-separated impurities, whose mutual distance is large compared to the quasiparticle coherence length, this modulation introduces position-dependent phase factors that preserve the symmetry structure of the QPI pattern, meaning the classification of the 
pairing symmetry (IR label) should remain robust. In the opposite limit of many aperiodically distributed impurities, the modulation approaches random speckle noise~\cite{KTB+25}, which can be handled by the self-supervised 
denoising preprocessing strategy discussed above. We therefore expect the classification of pairing symmetry to remain reliable, while the regression of continuous parameters such as $\Delta$ may be more sensitive to the additional amplitude modulation introduced by multiple impurities.

Another possibility is to have mixtures of inequivalent scatterers, whose different impurity species correspond to different values of the scalar impurity strength $v$. The QPI signal generated by a collection of species with strengths $v_1, v_2, \ldots, v_n$ is a superposition of contributions, each carrying the same symmetry structure, determined by the superconducting gap $\Delta_\mathbf{k}$ and the normal-state band structure, but rescaled by different amplitude factors. Since $v$ enters 
the T-matrix as an overall coupling strength and does not alter the symmetry properties of the Green's function, the symmetry-encoding features of the QPI pattern are preserved regardless of the mixture of scalar species present. We therefore expect the classification of the pairing symmetry to remain robust, while the regression of continuous parameters such as $\Delta$ may 
be more sensitive to the additional amplitude modulation. 

Finally, we comment on the effect of correlated spatial distributions of impurities, such as clustering or short-range order, which can introduce spurious scattering vectors into the QPI map. The classification of the pairing symmetry is expected to remain robust provided that the impurity distribution preserves, on average, the point group symmetry of the underlying lattice, since in this case the spurious scattering vectors respect the same symmetry as the intrinsic QPI features. When the impurity distribution breaks the lattice symmetry locally, the resulting spurious features can be suppressed by symmetrization of the QPI image, averaging over the symmetry operations of the underlying point group, which is a standard image processing step routinely applied in experimental QPI analysis to reduce noise and enhance contrast. This step is fully compatible with the self-supervised denoising strategy discussed above and can be applied jointly to experimental QPI images before feeding them into the classification network.

\section{Conclusions}\label{conclusions}
We introduced a machine-learning protocol that infers superconducting pairing symmetries and order parameters directly from simulated quasiparticle interference images of monolayer NbSe$_2$. By training a two-head convolutional neural network, we simultaneously identified the type of the superconducting pairing function, classified according to the irreducible representation of the ${\bf D}_{3{\rm h}}$ group, and continuous parameters of the superconducting state, including singlet–triplet mixing and the gap magnitude. Regularization through carefully chosen input augmentations yields models that remain robust under experiment-like noise conditions.

Our results show that the QPI-to-parameter inverse problem is tractable with high accuracy for most symmetry channels, even within a minimal scalar Anderson-impurity description and at a single excitation energy. At the same time, they expose intrinsic limitations of this framework: certain representations (notably 
$A_1^u$ and $A_2^u$) are effectively degenerate in their QPI response and are best treated as a merged class, and the time-reversal breaking nature of superconducting pairing functions transforming according to two-dimensional irreducible representations cannot be resolved within the scalar impurity model. 

These insights establish a concrete route toward symmetry- and parameter-resolved analysis of QPI data in unconventional superconductors and highlight clear extensions of the framework, such as incorporating magnetic and orbital-selective impurity channels, energy-resolved QPI, or more realistic tunneling matrix elements, to further lift degeneracies and enhance identifiability in multi-component pairing states.

\section*{Acknowledgments}
\acknowledgments
Research results was obtained using the computational resources procured in the national project National competence centre for high performance computing (project code: 311070AKF2) funded by European Regional Development Fund, EU Structural Funds Informatization of society, Operational Program Integrated Infrastructure. A.H. acknowledges the financial support provided by the Ministry of Education, Research, Development and Youth of the Slovak Republic, provided under Grant numbers APVV-21-0272 and VEGA 2/0133/25.
M.G.~acknowledges financial support provided by the Ministry of Education, Research, Development and Youth of the Slovak Republic, provided under Grant No. VEGA 1/0104/25 and the Slovak Academy of Sciences project IMPULZ IM-2021-42, and support of the QM4ST project funded by Programme Johannes Amos Commenius, call Excellent Research (Project No. CZ.02.01.01/00/22\_008/0004572).
M.M. acknowledges the financial support by the EU NextGenerationEU through the Recovery and Resilience Plan for Slovakia under the Project No. 09I02-03-V01-00012, by the APVV grant APVV-23-0430, and VEGA grants 2/0081/26 and 2/0133/25.
\section*{Data Availability Statement}
The quasiparticle interference datasets generated and analyzed during the current study (total size $\approx$ 50 GB) are not publicly archived due to their volume, but are available from the corresponding author on reasonable request.

\appendix
\section{Superconducting gap function}\label{appendixA}

Here, we provide a set of allowed superconducting gap function that can be constructed in a single-band approximation for a free-standing NbSe$_2$ monolayer, having a ${\bf D}_{3{\rm h}}$ symmetry.
The procedure for obtaining all symmetry allowed superconducting gap functions is equivalent to the case defined in~\cite{HMG24}, albeith different symetry group is imposed, the one of a freestanding NbSe$_2$ monolayer, not of NbSe$_2$ on a substrate.

Superconducting pairing function can be classified according to the IRs of the ${\bf D}_{3{\rm h}}$ group: four one-dimensional IRs $A_1^g$, $A_1^u$, $A_2^g$, $A_2^u$, and two two-dimensional IRs $E^g$ and $E^u$. Note that labels $g$ and $u$ represent parity quantum number with respect to the horizontal mirror plane symmetry $\sigma_{\rm h}$. 
As mentioned in the main text, it is additionally possible to classify superconducting pairing functions into one singlet (s) and two distinct triplet channels $(t,z)$ and $(t,xy)$. Each of the pairing channels are characterized by the  appropriate spin functions:
$d_0={\rm i}\sigma_y\sigma_0$ for singlet, $d_z={\rm i}\sigma_y\sigma_z$ for $(t,z)$ and $(d_x,d_y)=({\rm i}\sigma_y\sigma_x,{\rm i}\sigma_y\sigma_y)$ for $(t,xy)$ spin triplet channel. 

Whereas in the case of 1D IRs superconducting pairing function is uniquelly determined by a single function, we will call it $\Delta_{A_{1/2}^{g/u},{\bf k}}^{(s)/(t,z)/(t,xy)}$,
 in the case of 2D IRs $E^{g/u}$, the general form of the gap function can be written as $\Delta_{E^{g/u},{\bf k}}^{c}=(\cos{\theta_{c}}\Delta_{E^{g/u}_1,{\bf k}}^{c}+{\rm e}^{{\rm i}\varphi_{c}}\sin{\theta_{c}}\Delta_{E^{g/u}_2,{\bf k}}^{c})$, $c=(s),(t,z),(t,xy)$, where 
$\Delta_{E^{g/u}_{1/2},{\bf k}}^{c}$ are real functions to be constructed using the symmetry-based procedure, where $\theta_c$ measures the size of the mixing between the two components, while the parameter ${\varphi_c\neq0}$ signals the time-reversal symmetry breaking nature of the gap function.

In the singlet case, classified using the single spinor function $d_0$,
nonzero gap functions belong to the IRs $A_1^g$, $A_2^g$, and $E^g$ and their explicit ${\bf k}$-dependence is the following 
\begin{eqnarray}\label{s}
\Delta_{A_1^g,{\bf k}}^{(s)}&=&\Big[2\cos{(k_x a)}+4\cos{\frac{k_x a}2}\cos{\frac{\sqrt{3}k_y a}2}\Big]d_0,\nonumber\\
\Delta_{A_2^g,{\bf k}}^{(s)}&=&2\Big[
\sin{\frac{k_xa}2}\sin{\frac{3\sqrt{3}k_ya}2}-
\sin{(2k_xa)}\nonumber\\
&&\times\sin{(\sqrt{3}k_ya)}+\sin{\frac{5k_xa}2}\sin{\frac{\sqrt{3}k_ya}2}\Big] d_0,\nonumber\\
\Delta_{E^g_{1},{\bf k}}^{(s)}&=&\Big[\cos{(k_xa)}-\cos{\frac{k_xa}{2}}\cos{\frac{\sqrt{3}k_ya}{2}}\Big]d_0,\nonumber\\
\Delta_{E^g_{2},{\bf k}}^{(s)}&=&\Big[\sqrt{3}\sin{\frac{k_xa}{2}}\sin{\frac{\sqrt{3}k_ya}{2}}
\Big] d_0.
\end{eqnarray}
We note that it is not possible to construct singlet gaps that transform according to IRs $A_1^u$, $A_2^u$, and $E^u$ due to symmetry constraints (in the case of the single-band approximation employed here, see Section~\ref{BdGsubsection}).

Secondly, we present the explicit formulas for the triplet gap functions constructed using the $d_z$ spinor function, classified in terms of the IRs 
$A_1^g$, $A_2^g$, and $E^g$ (due to symmetry constraints, for the same reason as in the singlet case, it is not possible to construct triplet gap functions with $d_z$ spinor that would belong to IRs $A_1^u$, $A_2^u$, and $E^u$) 
\begin{eqnarray}\label{tz}
\Delta_{A_1^g,{\bf k}}^{(t,z)}&=&\Big[\Big(\cos{\frac{k_xa}{2}}-\cos{\frac{\sqrt{3}k_y a}{2}}\Big)\sin{\frac{k_xa}2}\Big]d_z,\nonumber\\
\Delta_{A_2^g,{\bf k}}^{(t,z)}&=&\Big[\sin{(\sqrt{3}k_ya)}-2\sin{\frac{\sqrt{3}k_ya}{2}}
\cos{\frac{3k_xa}{2}}\Big]d_z, \nonumber\\
\Delta_{E^g_{1},{\bf k}}^{(t,z)}&=&\Big[
\sin{(k_x a)}+\sin{\frac{k_xa}2}\cos{\frac{\sqrt{3}k_y a}2}\Big] d_z,\nonumber\\
\Delta_{E^g_{2},{\bf k}}^{(t,z)}&=&\Big[\sqrt{3}\cos{\frac{k_xa}2
\sin{\frac{\sqrt{3}k_ya}2}}\Big] d_z.
\end{eqnarray}
Again, in the case of the 2D representations the general form of the gap function can be written as $\Delta_{E^u,{\bf k}}^{(t,z)}=(\cos{\theta_t}\Delta_{E^u_1,{\bf k}}^{(t,z)}+{\rm e}^{{\rm i}\varphi_t}\sin{\theta_t}\Delta_{E^u_2,{\bf k}}^{(t,z)})$, with $\theta_t/\varphi_t$ measuring the component mixing/time-reversal symmetry breaking.  

Finally, we present supreconducting gap function constructed using the multiplet $(d_x,d_y)=({\rm i}\sigma_y\sigma_x,{\rm i}\sigma_y\sigma_y)$
\begin{eqnarray}\label{txy}
\Delta_{A_1^u,{\bf k}}^{(t,xy)}&=&\Big[ 3\cos{\frac{k_x a}2} \sin{\frac{\sqrt{3} k_ya}2}\Big] d_x-\Big[\sqrt{3}\Big(2\cos{\frac{k_xa}2}\nonumber\\
&&+\cos{\frac{\sqrt{3}k_y a}2}\Big)\sin{\frac{k_xa}2}\Big] d_y,\nonumber\\
\Delta_{A_2^u,{\bf k}}^{(t,xy)}&=&\Big[\Big(2\cos{\frac{k_xa}2}+\cos{\frac{\sqrt{3}k_y a}2}\Big)\sin{\frac{k_xa}2}\Big] d_x\nonumber\\
&&+\Big[\sqrt{3}\cos{\frac{k_x a}2} \sin{\frac{\sqrt{3} k_ya}2}\Big] d_y,\nonumber\\
\Delta_{E^u_{1},{\bf k}}^{(t,xy)}&=&\Big[2\sin{(\sqrt{3}k_y a)}-\cos{\frac{3k_x a}2}\sin{\frac{\sqrt{3}k_ya}2}\Big] d_x\nonumber\\
&&+\Big[\sqrt{3}\sin{\frac{3k_xa}2}\cos{\frac{\sqrt{3}k_ya}2}\Big] d_y,\nonumber\\
\Delta_{E^u_{2},{\bf k}}^{(t,xy)}&=&\Big[\sqrt{3}\sin{\frac{3k_x a}2}\cos{\frac{\sqrt{3}k_y a}{2}}\Big] d_x\nonumber\\
&&-\Big[3\cos{\frac{3k_x a}2}\sin{\frac{\sqrt{3}k_y a}2}\Big] d_y,
\end{eqnarray}
classified according to the IRs $A_1^u$, $A_2^u$, and $E^u$; due to symmetry constraints, no gap functions that transform according to IRs $A_1^g$, $A_2^g$, and $E^g$ can be constructed.

In previous equations, Eqs.~\eqref{s}-\eqref{txy}, we have presented all possible superconducting gaps, classified in terms of IRs of the given ${\bf D}_{3{\rm h}}$ group. An additionalnal feature of this classification is the possibility to have
mixed singlet-triplet components, since in the $A_1^g$, $A_2^g$, and $E^g$ case, both singlet $(s)$ and triplet $(t,z)$ components are present. Taking this into account, the most general form of the gap function that transform according to 1D IRs
\begin{eqnarray}\label{1Dgaps}
\Delta_{A_1^u,{\bf k}}&=&\Delta\Delta_{A_1^u,{\bf k}}^{(t,xy)},\nonumber\\
\Delta_{A_2^u,{\bf k}}&=&\Delta\Delta_{A_2^u,{\bf k}}^{(t,xy)},\nonumber\\
\Delta_{A_1^g,{\bf k}}&=&\Delta(\cos{\theta}\Delta_{A_1^g,{\bf k}}^{(s)}+\sin{\theta}\Delta_{A_1^g,{\bf k}}^{(t,z)}),\nonumber\\
\Delta_{A_2^g,{\bf k}}&=&\Delta(\cos{\theta}\Delta_{A_2^g,{\bf k}}^{(s)}+\sin{\theta}\Delta_{A_2^g,{\bf k}}^{(t,z)}),
\end{eqnarray}
whereas in the 2D case the mose general form is equal to
\begin{eqnarray}
\Delta_{E^u,{\bf k}}&=&\Delta(\cos{\theta_1}\Delta_{E^u_1,{\bf k}}^{(t,xy)}+{\rm e}^{{\rm i}\varphi_1}\sin{\theta_1}\Delta_{E^u_2,{\bf k}}^{(t,xy)}),\label{EuAPP}\\
\Delta_{E^g,{\bf k}}&=&\Delta\Big[\cos{\theta}
(\cos{\theta_s}\Delta_{E^g_1,{\bf k}}^{(s)}+{\rm e}^{{\rm i}\varphi_s}\sin{\theta_s}\Delta_{E^g_2,{\bf k}}^{(s)})+\nonumber\\
&&\sin{\theta}(\cos{\theta_t}\Delta_{E^g_1,{\bf k}}^{(t,z)}+{\rm e}^{{\rm i}\varphi_t}\sin{\theta_t}\Delta_{E^g_2,{\bf k}}^{(t,z)})\Big],\label{EgAPP}
\end{eqnarray}
where $\Delta$ represents the amplitude of the superconducting gap. As discussed above, in the case of the 1D IRs $A_1^u$ and $A_2^u$ no mixing occurs, whereas in the case of 1D IRs $A_1^g$ and $A_2^g$ there is only one mixing parameter. In the case of the 2D IRs the situation becomes more complicated since in the $E^u$ case, there are two parameters included in the mixing of the two components of the same gap constructed using the $d_{xy}$ triplet. In the second case, IR $E^g$, there is a mixing between the singlet and the triplet component $d_z$. The total number of parameters in this case is 5, where $\theta$ represents the mixing between the singlet and the triplet component, whereas $(\theta_s,\varphi_s)/(\theta_t,\varphi_t)$ are the mixing parameters within the 2D space of the singlet/$d_z$ space.

In the main text, we focus on all 1D IRs $A_1^u$, $A_1^g$, $A_2^u$, and $A_2^g$~\eqref{1Dgaps}. To simplify our study and to prove our concept that machine learning can be used to distinguish different features of the superconductivity, we focus on the singlet component of the 2D IR $E^g$ only (see Eq.~\eqref{2Dsinglet}), whereas keeping $\Delta_{E^u,{\bf k}}$ as in~Eq.~\eqref{EuAPP}. 

\section{Architecture of the neural network}\label{appendix:nn_arch}
\begin{figure*}[t]
    \centering \includegraphics[width=0.6\linewidth]{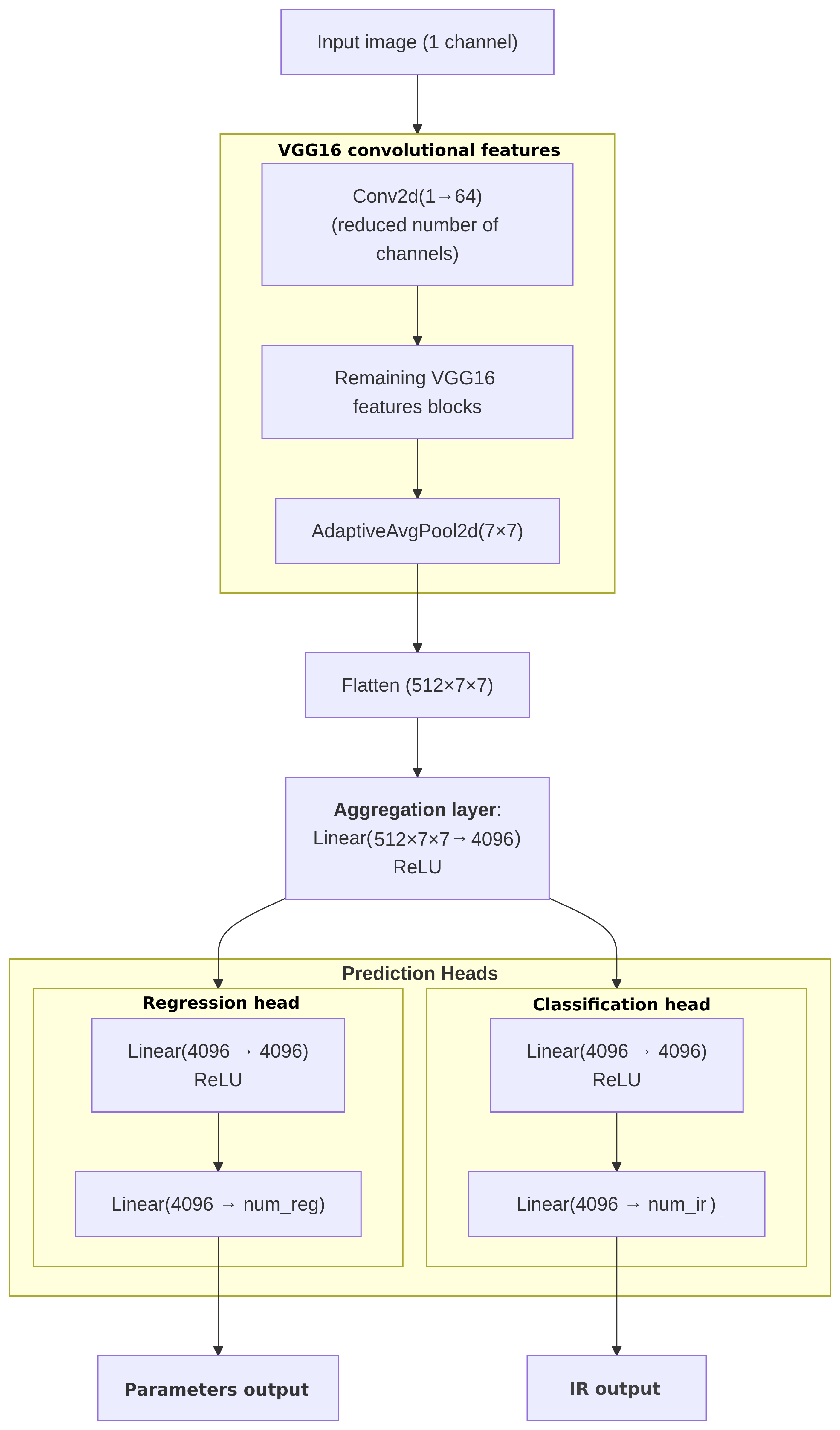}
    \caption{{\it Architecture of the neural network.} The model is based on VGG16 ~\cite{SZ15,DZM+21} architecture, input channels reduced from 3 to 1, prediction head replaced with two custom prediction heads, one for classification (IRs), the other one for regression (parameters).}
    \label{fig:nn_architecture}
\end{figure*}

Here, we provide schematics of the neural network architecture (see Fig.~\ref{fig:nn_architecture}) used to predict the type of superconducting pairing and the microscopic parameters of the BdG Hamiltonian.

\section*{References}
\bibliography{bibliography.bib}
\end{document}